\documentclass[amsmath,amssymb,reprint,aps,pra,floatfix]{revtex4-2}
\usepackage{graphicx}
\usepackage{lineno}

\usepackage[utf8]{inputenc}
\usepackage{siunitx}
\DeclareSIUnit{\sqrthertz}{\sqrt{\unit{\hertz}}}
\usepackage[german, english]{babel}

\begin{document}

\title{Optical steering of a large ring laser}

\author{Jannik Zenner}
  \affiliation{Physikalisches Institut, Universität Bonn, Nussallee 12, 53115 Bonn, Germany}
\author{Karl Ulrich Schreiber}
  \affiliation{Research Unit Satellite Geodesy, Technical University of Munich, 80333 Munich, Germany}
  \affiliation{School of Physical Sciences, University of Canterbury, Christchurch 8140, New Zealand}
\author{Simon Stellmer}
  \email{stellmer@uni-bonn.de}
  \affiliation{Physikalisches Institut, Universität Bonn, Nussallee 12, 53115 Bonn, Germany}

\date{\today}

\begin{abstract}

A common approach to reduce the linewidth of a laser is an increase of its resonator length. In large gas lasers, however, the frequency spacing between longitudinal modes of the resonator easily becomes significantly smaller than the Doppler-broadened width of the gain profile. As a consequence, the laser might operate on a multitude of modes simultaneously, or jump between modes. Such unstable operation cannot be tolerated in metrological or sensing applications, such as ring laser gyroscopes. 
Here, we propose and demonstrate a method to establish stable operation on a chosen mode index by optically steering the ring laser to a desired mode index through injection locking with an external laser. The injected mode reliably follows the external steering. Intra-cavity backscattering can even cause the counter-propagating, non-injected mode to follow the external steering as well.
\end{abstract}

\maketitle

\section{Introduction}

Injection locking of lasers \cite{Stover1966} is a well-established technique and often used in a so-called master/slave configuration, where a "master" laser (often narrow-linewidth and well-controlled in frequency) provides the seed for a much more powerful “slave” laser. This second laser then follows the phase and frequency of the seed laser. This method is widely used in solid-state lasers but has, to our knowledge, only rarely been applied to gas lasers. Examples include Ar ion lasers \cite{Man1984} and He-Ne lasers \cite{Stover1966, Dunn1983}, where two lasers of the same kind have been coupled. \\
Large He-Ne ring lasers \cite{Schreiber2023, Schreiber2025, DiVirgilio2019, Gebauer2020} may feature a perimeter $P$ larger than \SI{10}{\meter}, corresponding to a free spectral range $f_{\text{FSR}} = c/P$ of less than \SI{30}{\mega \hertz}. For a commonly used mixture of \SI{10}{\milli \bar} of Helium and \SI{0.2}{mbar} of Neon, the width of the gain profile in He-Ne lasers at a lasing frequency $f_{\text{L}}$ is derived in Refs.~\cite{Bennett1962, Graham2010}, determined by the thermal Doppler profile of the neon gas \cite{Siegman1986}:
\begin{align}
    \label{align:doppler}
    \Delta f_{\text{FWHM}}= \frac{f_{\text{L}}}{c} \sqrt{\frac{8 \ln(2) k_{\text{B}}T}{m}}.
\end{align}
The gas temperature of the gain medium can be assumed to be $T=$ \SI{300}{\kelvin} for the low radio-frequency (RF) power of around \SI{1.5}{\watt} used for plasma excitation in sensor applications \cite{Graham2010}. Further, an average atomic mass of $m=$ \SI{21}{\atomicmassunit} is assumed for the commonly used 50/50 mixture of the $^{20}$Ne and $^{22}$Ne isotopes. The Doppler-broadened width of the gain curve then amounts to \SI{1.3}{\giga \hertz}. Near its maximum, the gain curve is essentially flat, and the laser operates multi-mode on a stochastic composition of longitudinal mode indices near the gain maximum. Drastic reduction of the driving power to just above the lasing threshold (“gain starvation”) will enforce single-mode operation, but the mode index will be chosen randomly from the mode indices within about \SI{50}{\mega \hertz} around the gain maximum. If not sufficiently stable, the laser might jump between different modes, run multi-mode, or operate on different mode indices for the two counter-propagating modes (“split mode”) \cite{Brotzer2025}. \\
The commonly used procedure to establish single-mode operation on a specific mode index is based on cycles of a deliberate increase and slow decrease in driving power, which are repeated until both counter-propagating modes – by chance – operate on the same desired mode index (“common mode”). A single such cycle may take minutes and may be required multiple times per hour, which is unsuitable if 100\% uptime is required. As an example, this method  renders only around 85\% uptime for the ROMY ring laser \cite{Brotzer2025}. In addition, variations in driving power and circulating optical power induce thermal instabilities that degrade performance.\\
In this Letter, we present a purely optical approach in which an external diode laser is injected into the large ring resonator of a He-Ne laser at a given mode index, forcing the He-Ne gas laser to jump to this mode as well. The injection laser is turned on only intermittently, and the gas laser follows suit on a timescale of milliseconds, given by the cavity decay time. The counter-propagating mode, important for gyroscope operation in the Sagnac scheme, may follow the injection laser through intra-cavity backscatter coupling \cite{Hurst2014} after a phase of split-mode operation. Importantly and different from common injection-locking schemes, the external light is injected only temporarily, and sensing in gyroscope operation is performed without injection light present.

\section{Setup}

\begin{figure}[t]
    \includegraphics[width=\columnwidth]{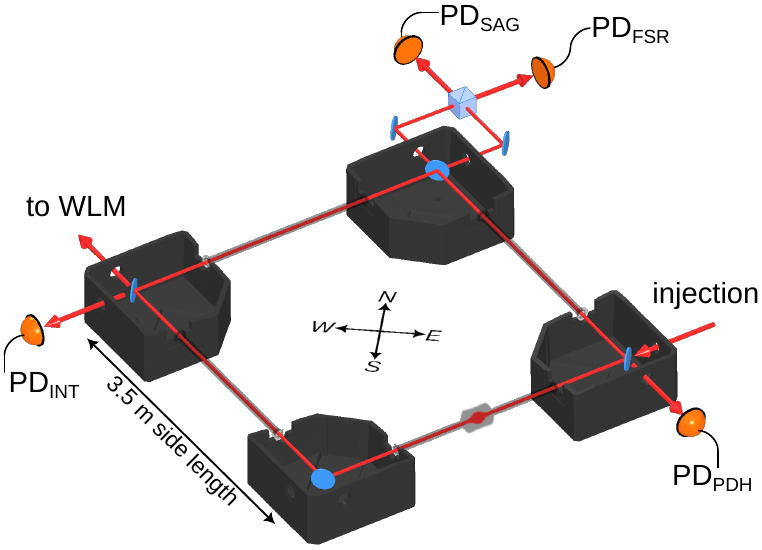}
    \caption{\label{fig:setup} Schematic view of the \SI{14}{\meter} perimeter ring laser cavity. The Sagnac and $f_{\text{FSR}}$ beats are measured with different photodiodes at the north corner. The external laser is injected in clockwise direction at the east corner. At the west corner, the clockwise transmission is coupled into a fiber to be evaluated by a wavelength meter and the counter-clockwise transmission is used to stabilize the lasing intensity of the active ring laser.}
\end{figure}

The setup, as shown schematically in Fig.~\ref{fig:setup}, is based on a square ring laser cavity of \SI{3.5}{\meter} side length inside a vacuum chamber, described in detail elsewhere \cite{Schreiber2009,Schreiber2006}. The resonator is formed by four mirrors with a radius of curvature of \SI{4}{\meter} and a finesse of about $F=$~36,000. The free spectral range of the cavity is $f_{\text{FSR}} = c/P =$ \SI{21.42}{\mega \hertz}, equating to a cavity linewidth $f_{\text{FWHM}} = f_{\text{FSR}}/F$ of about \SI{600}{\hertz}. The ring laser is operated in s-polarization. The penetration depth into the mirror coatings is polarization-dependent, causing small differences in $P$ between s- and p-polarization that split the respective resonances by many MHz. \\
When operated as an active ring laser gyroscope, as described in Ref.~\cite{Zenner2025}, the lasing power is about \SI{90}{\nano \watt}, which is detected on the PD$_{\text{INT}}$ photodiode for the counter-clockwise direction's beam at the west corner. This signal is fed into a PID controller that acts via a feedback loop on the power of the radio frequency drive that excites the plasma, keeping the lasing intensity in this direction constant.\\
Both, the clockwise and counter-clockwise transmissions are combined with a non-polarizing beam splitter cube at the north corner. If the lasing occurs on common modes, the \SI{50}{\kilo \hertz} bandwidth PD$_{\text{SAG}}$ photodiode records a Sagnac beat frequency $\delta f$ that is proportional to the sensor's rotation rate $\Omega$ according to
\begin{align}
    \label{align:sag}
    \delta f = \frac{4 A}{\lambda P} \Omega \sin\theta,
\end{align}
where $A$ is the area, $\lambda$ is the wavelength of the light and $\theta$ is the projection angle of the surface vector onto the rotation vector, i.e.~the latitude of $\theta = 50^{\circ}$ 43' 41.9'' N. The other output of the beam combiner is recorded with a fast \SI{400}{\mega \hertz} bandwidth photodiode PD$_{\text{FSR}}$, which allows observation of a beat frequency at multiples of $f_{\text{FSR}}$, if the ring laser is operating in split mode.\\
An external diode laser (TOPTICA DL PRO with an unlocked linewidth of about \SI{100}{\kilo \hertz}) is tuned to the same frequency as the active He-Ne lasing. It is frequency-modulated at \SI{5}{\mega \hertz} by an electro-optical modulator, frequency-shifted by an acusto-optical modulator (AOM), and passed through a polarization-maintaining fiber to be coupled into the ring cavity at the east corner, labelled "injection" in Fig.~\ref{fig:setup}. The reflection of these \SI{185}{\micro \watt} of light is detected with a \SI{150}{\mega \hertz} bandwidth photodetector PD$_{\text{PDH}}$ to perform a Pound-Drever-Hall lock \cite{Black2001} of the diode laser using a fast proportional-integral-derivative (PID) controller (TOPTICA FALC PRO). This stabilizes the laser frequency to the center of a cavity resonance, within about \SI{100}{\hertz}. As previously shown for this setup \cite{Zenner2026}, passive operation of the ring cavity, which is currently a research field of major interest \cite{Feng2023, Chen2025}, may occur independently from the active gas laser operation. Careful positioning of the cavity mirrors to within \SI{50}{\micro \meter} of a geometrical plane, beamwalking of the incoupling of the injection, and the intrinsic behavior of the \SI{4}{\milli \meter} diameter plasma cell tube as an aperture, all lead the active ring laser to be only operating on TEM$_{0,0}$ modes of the cavity. The same is true for the external “passive” operation.\\
The clockwise transmission at the west corner is coupled into a fiber that is connected to a HighFinesse WS8 wavelength meter (WLM), which enables continuous monitoring of the lasing frequency $f_{\text{L}}$ of the ring with \SI{1}{\mega \hertz} precision, at a rate of \SI{1.75}{\second} limited by the finite integration time of the wavelength meter.\\
The AOM is set up to shift the diode laser light by a constant frequency of \SI{200}{\mega \hertz} if enabled. It is used as a rapid switch to turn on and off the injection, as the PID feedback loop is set up to automatically lock the laser to the cavity resonance nearest to the laser's frequency, which is monitored at a second channel of the wavelength meter.

\section{Results}

\begin{figure}[t]
    \includegraphics[width=\columnwidth]{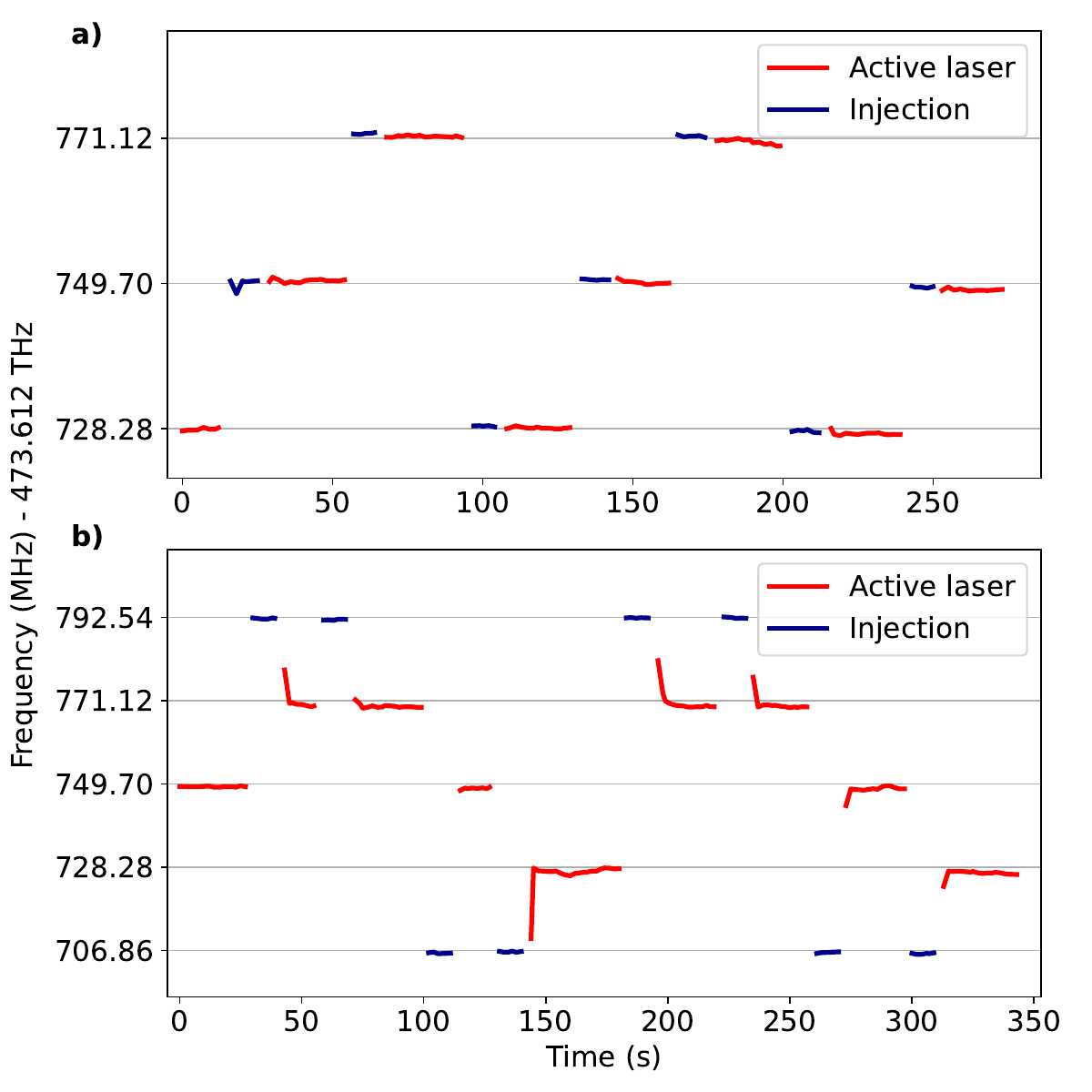}
    \caption{\label{fig:injection} Wavelength meter measurement of the active ring laser frequency (red) and the frequency of the diode laser beam, when it is injected into the ring cavity (blue). a) The active laser modes follow the frequency of the injected beam within $f_{0}\pm 1f_{\text{FSR}}$. b) The active laser does not follow injections at frequencies of $f_{0}\pm2f_{\text{FSR}}$, jumping to a frequency within $f_{0}\pm1f_{\text{FSR}}$.}
\end{figure}

\begin{figure*}[t]
    \includegraphics[width=\textwidth]{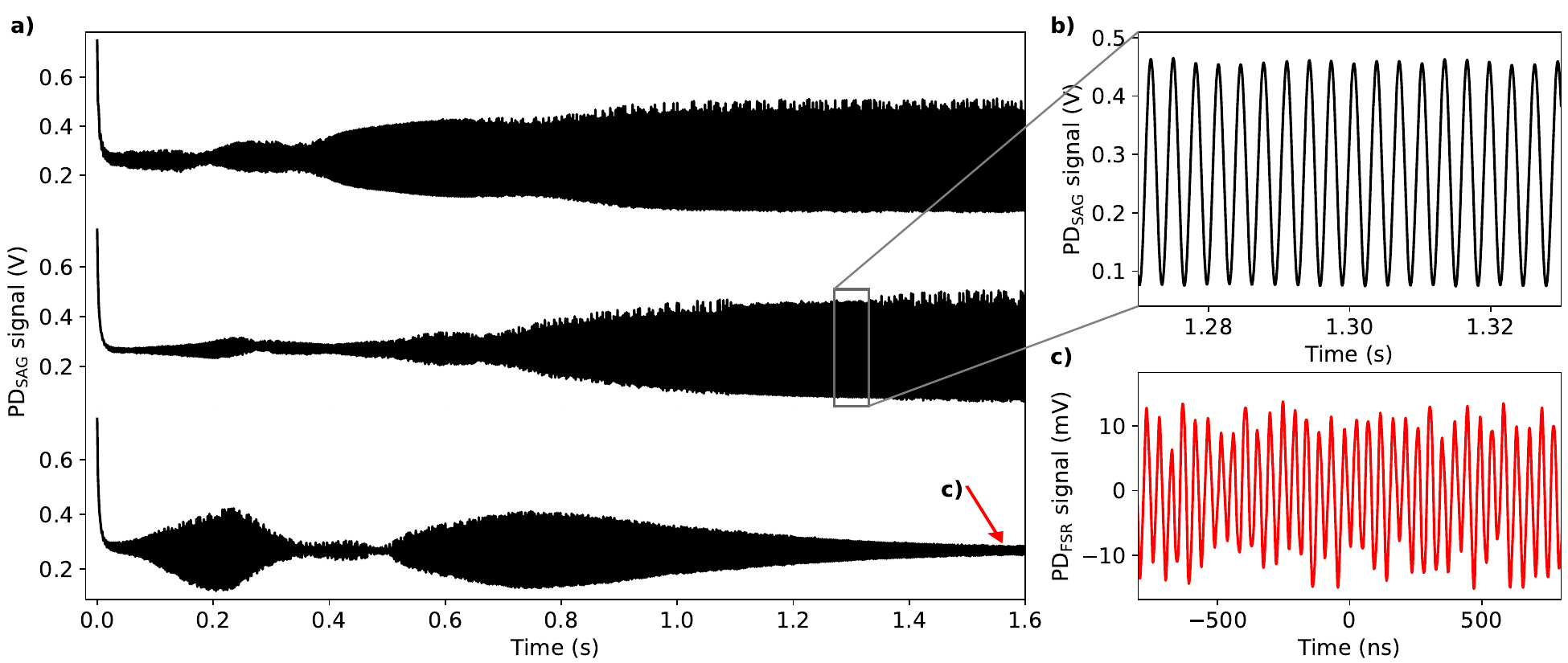}
    \caption{\label{fig:dynamics} a) Amplitude swelling in the PD$_{\text{SAG}}$ signal after injections. The first and second datasets show an intact Sagnac signal after about \SI{1}{\second}. The third dataset shows an injection that yields no Sagnac signal after mode competition dynamics over many seconds. b) Insert showing the Sagnac beat of about \SI{312}{\hertz}, observed during common mode operation. c) Example of the PD$_{\text{FSR}}$ signal showing the $f_{\text{FSR}} =$ \SI{21.42}{\mega \hertz} beat if split mode operation is present, as in the indicated position in a).}
\end{figure*}

The clockwise active lasing frequency $f_{\text{L}}$ is constantly monitored by the WLM and plotted in Fig.~\ref{fig:injection} (red data). While injections are performed, the red data shows gaps, because the additional light of the transmission of the external laser locked to the cavity causes the WLM to be overexposed during this period. The frequency of the external laser during the injection is also recorded, as shown by the blue graph. The data in  Fig.~\ref{fig:injection}~a) shows seven injections of about \SI{10}{\second} duration at the central frequency of $f_{0} =$ \SI{473.6127497}{\tera \hertz} or at adjacent modes at $f_{0}\pm 1f_{\text{FSR}}$. After each injection, from the next datapoint onward, it is clear that the ring laser mode always follows the stabilized external laser. For all trials performed within this study, the active mode \textit{always} followed the injection, and the probability of the active mode getting optically steered by the injection light can be assumed to be 100\%.\\
For this experiment, \SI{185}{\micro \watt} of light was sent onto the cavity with an incoupling efficiency of 11\%. To assess the dependence on injection strength, the light power was varied all the way down to \SI{18}{\micro \watt} by adjusting the RF power sent to the AOM. For the lowest injection strength, the injected light power circulating in the cavity was about twice the active power.  Even at this level, no phenomenological difference was found, showing robustness of this behavior even at low injection power. Further decrease of the injection power was not feasible without compromising the quality of the PDH lock in the current setup.\\
Furthermore, the success probability in dependence of the injection duration was investigated. All the way down to the sub-millisecond range, $f_{\text{L}}$ followed the injection every single time. Operation on the new mode index was observed to be stable over many hours, ultimately limited by mode jumps induced by length drifts of the cavity.\\

We now explore the limits of our steering method in terms of frequency. The same injection procedure is performed, but this time the frequency of the external laser is set to $f_{0} \pm 2f_{\text{FSR}}$, as shown in Fig.~\ref{fig:injection}~b). Four trials at $f_{0} + 2f_{\text{FSR}}$ and four at $f_{0} - 2f_{\text{FSR}}$ are shown, where, unlike above, the active ring does \textit{not} follow the frequency of the injected light. Instead, $f_{\text{L}}$ jumps to a seemingly random frequency within $f_{0}\pm~1f_{\text{FSR}}$, on a timescale faster than the \SI{1.75}{\second} sample rate of the WLM. Some injection attempts show the first point, after the WLM being overexposed during the injection process, to be skewed towards the injection frequency. This might indicate that the active laser follows the injection only for a fraction of the WLM sample rate period, before jumping to a mode closer to $f_{0}$. Without further investigation, we conclude that the frequency interval, in which the laser can be steered reliably by external injection, is between $2f_{\text{FSR}}=$ \SI{42.84}{\mega \hertz} and $4f_{\text{FSR}}=$ \SI{85.68}{\mega \hertz} for this given setup.\\

During injection, the Sagnac photodiode PD$_{\text{SAG}}$ records a transmission signal of \SI{4}{\volt}, compared to the \SI{0.5}{\volt} amplitude of the Sagnac beat of the active operation. When the injection is turned off, by switching off the radio frequency to the AOM, PD$_{\text{SAG}}$ records an exponential ringdown decay of the injection light intensity with the same ringdown time of $\tau =$ \SI{270}{\micro \second} as measured when turning off the active ring laser. Fig.~\ref{fig:dynamics} a) shows the tail of this ringdown decay for three examples where the ring laser was injected from $f_{0}$ to the $f_{0} + 1f_{\text{FSR}}$ mode. Observing the time after an injection, it can be seen that a Sagnac beat is not immediately present but might build up on a timescale of about \SI{1}{\second}.\\
Following the observations explained above, we expect the clockwise mode to be operating solely on the single injected mode at this time. The dynamics observed here are purely caused by the non-injected direction, i.e., the counter-clockwise mode. Driven by plasma-enhanced backscatter coupling, the counter-clockwise mode partially starts lasing on the injected mode index as well. It enters a multi mode state between the injected and the previous mode, where mode competition happens on a timescale of seconds. The ratio of the common and split mode lasing oscillates during this multi mode phase, as apparent from variations of the Sagnac amplitude. This multi mode competition collapses to a single mode after about \SI{0.8}{\second} to \SI{1.5}{\second} onto either common mode operation, as seen in the first two shown examples, or onto a split mode operation, as shown in the third. Fig.~\ref{fig:dynamics} b) shows a zoom into the clean Sagnac signal after common mode operation has been established.\\

When split mode operation is present, the PD$_{\text{FSR}}$ signal shows a beat at the frequency $f_{\text{FSR}}$ as seen in Fig.~\ref{fig:dynamics} c). The amplitude modulation is likely caused by unstable mode competition present in a split mode configuration, indicating the impracticality of this operation regime for a sensitive gyroscope sensor. This mode competition arises from backscatter coupling, which causes a fraction of the light to propagate on the same mode as the counter-propagating direction, thereby also producing a small residual Sagnac signal at this position.\\
It is interesting to note that this observed beat always occurs at a frequency of $1f_{\text{FSR}}$, and a split mode between two non-adjacent modes is never observed, not even when injecting the ring laser from $f_{0}-1f_{\text{FSR}}$ to $f_{0}+1f_{\text{FSR}}$ or vice versa.\\
\\
Further, we observe that the backscatter-induced mode competition after an injection causes the non-injected direction to mode jump in about two thirds of all cases, while the injected direction always follows the injected frequency, as seen in Fig.~\ref{fig:injection} a). The intricate process of backscatter coupling is explained in Ref.~\cite{Hurst2014}, and more work is required to elucidate the origin of the asymmetry. Still, it is interesting to observe that the injected mode is more robust than the counter-propagating mode, even on timescales many orders of magnitude longer than $\tau$.

\section{Conclusion}

In this Letter, we have presented a method to optically steer the longitudinal mode index of a ring laser. Injecting an external laser proves to be, even with low power and at a duration as short as the ringdown time, a robust way to enforce lasing at one of three neighboring longitudinal cavity modes with a spacing of $1\,f_{\text{FSR}}=$ \SI{21.42}{\mega \hertz}. The counter-propagating mode does not disturb this steering and may, via backscatter-induced interaction, settle on the mode index of the injected direction.\\
As the steering of the injected direction is always successful, we suspect that an extended scheme that includes a second branch of the laser, shifted by $\delta f$ and injected in the counter-propagating direction, could steer both directions at the same time. If no distinct challenges emerge for this simultaneous bi-directional steering, it can be of great technological value for large ring lasers to mitigate split-mode operation and thus enable proper Sagnac operation at close to 100\% uptime.\\
This method becomes increasingly important for ring lasers of increased size. The perimeter of ring lasers can exceed \SI{100}{\meter}: the UG2 ring laser, for example, had a free spectral range of \SI{2.5}{\mega \hertz}, and would run on dozens of mode indices \cite{Hurst2009}.

\section*{Acknowledgments}
We are indebted to U.~Hugentobler and the Forschungseinrichtung Satellitengeodäsie at TU Munich for the loan of the ring laser hardware. We acknowledge fruitful discussions with Th.~Gereons, J.~Kodet, H.~Igel, and A.~Brotzer, as well as experimental support from P.~Hänisch. \\

\section*{Data Availability Statement}
The data that support the findings of this study are available from the corresponding author upon reasonable request.

\section*{Disclosures}
The authors declare no conflicts of interest.\\

\section*{Funding}
We acknowledge financial support from the European Research Council ERC through grants No.~101123334 and 101213032.

\bibliography{bib}

\end{document}